\documentclass[fleqn,12pt,twoside]{article}
\usepackage[headings]{espcrc1}

\readRCS
$Id: espcrc1.tex,v 1.2 2004/02/24 11:22:11 spepping Exp $
\usepackage{graphicx}
\usepackage[figuresright]{rotating}

\newcommand{\AmS}{{\protect\the\hboxfont2
  A\kern-.1667em\lower.5ex\hbox{M}\kern-.125emS}}

\title{Shedding Light on the Symmetries of Dark Matter}
       
\author{Susan Gardner \address[UK]{Department of Physics and Astronomy, 
University of Kentucky, Lexington, KY 40506-0055 USA}%
         \thanks{I acknowledgement partial support from the U.S. Department of Energy
under contract DE--FG02--96ER40989, and I thank David C. Latimer for his 
collaboration on the topic of dark-matter-mediated effects on the propagation of light.}}

\runtitle{Symmetries of Dark Matter}
\runauthor{S. Gardner}

\begin{document}

\maketitle

\begin{abstract}
I consider symmetries which could explain observed properties of 
dark matter, namely, its stability on Gyr time scales or its relic density 
and discuss how such symmetries can be discovered through the study of
the propagation and polarization of light in its transit through dark matter. 
\end{abstract}

\section{Preamble}

The existence of dark matter is established from 
galactic to cosmological distance scales if gravity is understood. 
The nature of dark matter is unknown, but we do know most of it must be 
stable or effectively so on Gyr time scales, 
not ``hot'' if it is a thermal relic, i.e., not relativistic at the 
time it decoupled from ordinary matter in the cooling early
Universe, and sufficiently weakly interacting that 
it possesses no substantial strong or electromagnetic charge. 
As yet unknown symmetries in the dark sector could explain these features. 
In this contribution, based, in part, on work performed in collaboration
with David C. Latimer~\cite{svgdcl}, I discuss direct detection schemes~\cite{svgdcl,svgprl,svgprd} 
which can establish their existence. 

The Standard Model provides no suitable dark-matter candidate, but theories 
which resolve the hierarchy problem to make the weak scale technically natural 
can. In the Minimally Supersymmetric Standard Model (MSSM), e.g., 
the dark-matter candidate is a Weakly Interacting Massive Particle (WIMP). 
Although the stability requirement is added to the MSSM through the imposition
of a discrete symmetry, a WIMP with a mass of ${\cal O}(100 \,\hbox{GeV})$ 
is compatible with the observed dark-matter density. However, it is also possible to reproduce it 
with lighter particles which possess stronger, i.e., weak but not of $G_F$ strength, 
mutual interactions~\cite{fengkumar}. 

If dark matter is not made of WIMPs, its stability need not be guaranteed 
by a discrete symmetry, and its relic density need not be fixed by thermal 
freezeout. What mechanisms then are operative and how do we discover
them? Its stability may be guaranteed by a hidden gauge symmetry. 
E.g., dark matter can possess a hidden U(1) symmetry. If the
gauge mediator is massless, although this is not a necessary condition, dark matter 
can have a {\em millicharge}~\cite{holdom,feldman}. 
If we determine that dark matter has a millicharge, we establish that
dark matter is stable by dint of a gauge symmetry, much as the electron is stable --- it cannot
decay and conserve its electric charge. 
We can discover a dark-matter millicharge from the appearance of dispersive
effects in the speed of light, and we shall discuss how the light curves of Gamma-Ray Bursts
(GRBs) can be used to this purpose~\cite{svgdcl}. 
Moreover, its relic density may not be a numerical accident; it could 
be related to the fraction of the cosmological energy density in baryons 
$\Omega_B$~\cite{technidark,diracdark}. If so, dark matter ought be asymmetric
in its particle-antiparticle content. 
The dark matter of such models is built of Dirac particles and can thus possess a magnetic moment. 
We can discover this through use of the gyromagnetic Faraday effect~\cite{svgprl,svgprd}
and can indeed
establish asymmetric dark matter, as we shall discuss. 

\section{Dispersive Effects in Light Propagation}

To discover the nature of dark matter we must probe its 
couplings to known matter. E.g., in many models, dark matter
can annihilate to photons and leptons, and recently much attention
has been paid to the possibility of indirect dark-matter detection 
via ``anomalous'' lepton excesses in high-energy cosmic
ray data. Models which yield annihilation of a dark-matter particle
$\chi$ to photons through $\chi\chi \to \gamma \gamma$ also produce, by crossing, 
a non-zero forward Compton amplitude and thus predicate 
an index of refraction $n(\omega)$ which can deviate from unity. 
This, in turn, can yield energy-dependent, or dispersive,
effects in light propagation. We can thus study the light
curves of GRBs with cosmological distance to hunt for dark matter directly
--- we refer to Ref.~\cite{svgdcl} for all details. 

We analyze the GRB data in a model-independent way by employing an effective
theory analysis. 
To realize this, we suppose that the 
photon energy $\omega$ is small compared to 
$\omega_{\rm th}$, the threshold 
energy required to materialize the particles 
to which the dark matter can couple. If dark matter is connected to weak-scale physics, 
then crudely $\omega_{\rm th} \sim {\cal O}(100 \,\hbox{GeV})$. 
We can then expand the forward Compton amplitude
in a power series in $\omega$ for $\omega \ll \omega_{\rm th}$; 
the symmetries of the forward Compton amplitude allow us to codify 
the terms which appear. 
Under Lorentz symmetry and $P$, $T$, and $C$ invariance, the forward Compton 
amplitude for $\gamma(k) + \chi(p)$ scattering in the dark-matter rest frame 
is~\cite{GGT54,G55,HHKK98}
\begin{equation} 
\mathcal{M}_r(k, p \to k,p) = f_1 (\omega) 
\mathbf{\epsilon'}^{\,\ast} \cdot \mathbf{\epsilon}
+ i f_2 (\omega) 
\mathbf{{\cal S}} \cdot \mathbf{\epsilon'}^{\,\ast} \times \mathbf{\epsilon} \,,
\end{equation}
where 
$\mathbf{\epsilon}$ ($\mathbf{\epsilon'}$) 
is the photon polarization in the initial (final) state and 
${\mathbf{\cal S}}$ is the dark-matter spin operator. 
The amplitude $\mathcal{M}_r(k, p \to k,p)$ is implicitly a $2\times2$ matrix 
in the photon polarization. 
Only its diagonal matrix elements describe dispersive effects in propagation, and thus
only $f_1$ matters. Under analyticity and unitarity, 
we have a dispersion relation for $f_1$~\cite{GGT54,G55}, where
\begin{equation}
\hbox{Re} f_1(\omega) - \hbox{Re} f_1(0)
 = \frac{4M\omega^2}{\pi} \int_0^\infty 
\mathrm{d}\omega'
\frac{\sigma(\omega')}{{\omega'}^2 - \omega^2}  \,,
\end{equation}
and the optical theorem has been used to replace $\hbox{Im} f_1(\omega)$ with 
the unpolarized cross section $\sigma$. The integral implicitly begins at
$\omega_{\rm th}$, so that for $\omega \ll \omega_{\rm th}$ we have finally 
\begin{equation}
\hbox{Re}\,n(\omega) = 1 + \frac{ \rho}{4M^2 \omega^2}\left( A_0 + A_2\omega^2 + \dots \right)\,,
\label{nparam}
\end{equation}
where $A_0 = \hbox{Re} f_1(0)$ and $A_i > 0$. Moreover, 
$\rho$ is the mass density and $M$ is the particle mass of the dark matter.
A low-energy theorem fixes $A_0=-2\varepsilon^2 e^2$ for dark matter of 
electric charge $\varepsilon e$~\cite{thirring}. Light emitted 
from a source at a distance $l$ from us possesses a 
frequency-dependent arrival 
time $t(\omega)$ after transit through dark matter: 
$t(\omega)= l(\tilde n + \omega \mathrm{d}\tilde n/\mathrm{d}\omega)$, 
where $\omega=k/\tilde n$ and $\tilde n \equiv \hbox{Re}\,n$. 
We must take the cosmological expansion into account as well~\cite{jacobpiran}, so that 
at red shift $z$ the photon energy is 
blue shifted by a factor of $1+z$ relative to its present-day 
value $\omega_0$. Thus 
\begin{equation}
\!t(\omega_0, z)= \!\!\int_0^z \!\frac{\mathrm{d}z'}{H(z')} 
\!\left(\! 1 + \frac{\rho_0(1+z')^3}{4 M^2}\!\left( \frac{-A_0}{ 
\!\left((1+z')\omega_0\right)^{2}} + A_2 + 3 A_4 (1 + z')^2 \omega_0^2 
+ ...
\!\right) \!\right) 
\end{equation}
with the Hubble rate $H(z') = H_0 \sqrt{(1+z')^3\Omega_M + \Omega_\Lambda}$, so that 
WMAP parameters characterize both the matter density and light travel time. 
We use the combined analysis of the WMAP five-year data and more 
as per Ref.~\cite{komatsu} in the $\Lambda$CDM cosmological model, namely, 
$H_0 = 70.5 \pm 1.3 \,\hbox{km\, s}^{-1}\hbox{Mpc}^{-1}$, whereas 
the fraction of the energy density
in matter relative to the critical density today is 
$\Omega_M = 0.274 \pm 0.015$ and 
the fraction of the energy density 
in the cosmological constant $\Lambda$ is $\Omega_\Lambda = 1 - \Omega_M$. 
Various strategies must be employed to isolate the $A_i$; here we focus on $A_0$, 
which is fixed by the dark-matter electric charge. 
Although the non-observation of 
frequency-dependent time lags in-vacuo 
from GRB data have been suggested as a means to limit the appearance of 
Lorentz violation~\cite{amelinocamelia}, the red-shift and frequency dependence
of the dark-matter and Lorentz violation scenarios are very different. 
We employ, however, the statistical analysis 
suggested in the latter context to separate propagation and GRB source effects~\cite{ellis2000}.

Gamma-Ray Bursts (GRBs) are very bright objects which are still 
appreciable at cosmological distances. 
Fermi expects to discover 200 per year~\cite{fermi}. 
GRBs possess several properties 
which correlate with their luminosity, so that they can be used to 
probe the Hubble diagram at large $z$ and to study the properties
of dark energy. A study of 69 GRBs extends the Hubble diagram to $z>6$ and is
consistent with the usual concordance model~\cite{sch07}, which supports
the use of the GRB data set in our current context. 
Fits to the various $A_i$ require observations of different energies; in particular, 
to constrain $A_0$ we require observations in the radio. To select the
GRBs to be included in our fit we demand that 
the energy of the GRB be compatible with the energy range 
of the Fermi Gamma-ray Burst Monitor (GBM) and 
that the radio flux detection be in the right location, be new, and 
be significantly non-zero. Moreover, we pick GRBs for which $z$ is measured. 
We find 53 GRBs in all to consider and include detected radio
frequencies of 75 GHz or less in our analysis. 
Our observable is $\tau = t(\omega_0^{\rm low},z) - t(\omega_0^{\rm high}, z)$, where 
$\omega \equiv \omega_0^{\rm low}$ henceforth. Thus we fit 
\begin{equation}
\frac{\tau}{1+z} = 
{\tilde A_0} \frac{K(z)}{\nu^2} + \delta((1+z)\nu) \,.
\label{fitfcn}
\end{equation}
Note the frequency $\nu\equiv\omega/2\pi$ and 
$K(z) \equiv (1+z)^{-1} \int_0^z \mathrm{d}z\,(1+z') H(z')^{-1}$, whereas 
$4\pi^2 \tilde A_0 = - A_0 \rho_0/4 M^2 = 
2\pi\alpha \varepsilon^2 \rho_0/M^2$ 
and $\rho_0\simeq 1.19\times 10^{-6}$ GeV/cm${}^3$~\cite{komatsu}. 
The function $\delta((1+z)\nu)$ allows for a frequency-dependent 
time lag for emission from the GRB in the GRB rest frame. To provide a context, 
we first consider the value of $|\varepsilon|/M$ which would result were we to attribute 
the time lag associated with the radio afterglow of one GRB to a propagation effect. 
Choosing the GRB with the largest value of $K(z)/\nu^2$, 
we have a time lag of $2.700\pm 0.006\,\hbox{day}$ associated with GRB 980703A at 
$z=0.967\pm 0.001$  measured at a frequency of $\nu=1.43\,\hbox{GHz}$. With 
Eq.~(\ref{fitfcn}), 
setting  $\delta=0$, and noting that $K(z)/\nu^2=1170\pm 10\,\hbox{Mpc}\,\hbox{GHz}^{-2}$ 
if the errors in its inputs are uncorrelated, the measured time lag fixes $|\varepsilon|/M\simeq 9\times
10^{-6}\,\hbox{eV}^{-1}$. Since there are no known examples of a radio afterglow preceding a
GRB, this one observation in itself represents a conservative limit. Turning to our fit, 
we include all observations in our GRB sample with frequencies of  $4.0- 75$ GHz in the GRB rest 
frame. A scale factor in the uncertainty 
in $\tau/(1+z)$ of 450 yields $\chi^2/\hbox{ndf}= 1.13$, 
with $\tilde A_0 = 0.0010 \pm 0.0019\,\hbox{day GHz}^{2}\, \hbox{Mpc}^{-1}$
and $\delta=0.65 \pm 0.10\, \hbox{day}$. 
Thus $\tilde A_0 < 0.005\,\hbox{day GHz}^{2}\, \hbox{Mpc}^{-1}$ 
at 95\% CL, and we determine 
\begin{equation}
\vert \varepsilon \vert/M <  1 \times 10^{-5} \hbox{ eV}^{-1}
\quad \hbox{at}\, 95\% \hbox{CL}\,,
\label{result}
\end{equation}
which is comparable to the limit derived from a single observation of GRB 980703A. 
Our fit uses radio observations at no less than 4 GHz in the GRB rest frame, so that 
the associated limit is operative if $\omega_{\rm th}/2\pi > 4\, \hbox{GHz}$, or, crudely, if 
$M > 8\times 10^{-6}\,\hbox{eV}$. 
We find a very large scale factor; this may stem, in part, 
from the circumburst environment~\cite{frail}.

We have found a direct observational limit on the 
electric-charge-to-mass ratio of dark matter. 
Millicharged matter limits also follow from 
the nonobservation of the effects of millicharged 
particle production. The strongest such bound from laboratory experiments
is $\vert \varepsilon \vert <  3-4 \times 10^{-7}$ 
for $M \le 0.05$ eV~\cite{ahlers}, so that for 
$M\sim 0.05\,\hbox{eV}$ the limits are crudely comparable. 
Indirect limits also emerge from stellar 
evolution constraints, for which the strongest is $|\varepsilon| < 2\times 10^{-14}$ for 
$M< 5\,\hbox{keV}$~\cite{davidson}, as well as from the manner in which numerical 
simulations of galactic structure confront observations~\cite{gradwohl,feng}. 
Such limits can be evaded; in some models, the 
dynamics which gives rise to millicharged matter are not operative at 
stellar temperatures~\cite{masso}; other models evade the galactic structure constraints~\cite{Qballs}. 
We estimate that our limit  would have to improve by 
${\cal O}(2\times10^{-3})$ before 
the contribution from ordinary charged matter, namely, from free electrons, could be apparent. 
One can expect linear improvement in the limit on $\varepsilon/M$
as $\nu$ decreases; the observation of prompt radio emission predicted
to exist at 30 MHz from GRBs, planned by the GASE collaboration~\cite{gase}, 
could yield considerably stronger limits.

\section{Gyromagnetic Faraday Rotation} 

Light in a medium with free magnetic moments can become 
{\it circularly birefringent} if the applied magnetic field 
$|\mathbf{B_{0}}|$ is non-zero. 
If $|\mathbf{B_{0}}|$ 
induces a magnetization, 
$\mathbf{{\cal M}}_{\rm 0}$, and the light is 
directed along the magnetic field, initially linearly polarized light can exhibit 
Faraday rotation after transit through the medium~\cite{polder}. 
In this case, the magnetic field associated with the propagating light wave induces
a component of the magnetization perpendicular to ${\mathbf {\cal M}_0}$; consequently, 
the wave vector in the medium depends on the helicity of the light. 
Referring to Refs.~\cite{svgprl,svgprd} for all details, we note, namely, that 
\begin{equation} 
k_\pm =\omega \sqrt{1 \pm \frac{\chi_0\omega_B}{\omega \pm \omega_B}} \,, 
\end{equation}
where $\chi_0 \equiv {\cal M}_0/B_0$, 
$\omega_B \equiv g\mu_M B_0$, and $\hbar=c=1$. 
The magnetic moment $\mu$ of a particle of mass $M$ is 
$\mu=S g \mu_M$ with $\mu_M\equiv e/2M$, where $S$ is its spin and $g$ is its
Land\'e factor. 
Expanding in  $\omega_B/\omega$, we find 
\begin{equation}
k_{\rm diff} \equiv  k_+ - k_- 
= \chi_0 \omega_B  +
\frac{\chi_0 \omega_B^3}{\omega^2} + 
\frac{\chi_0^2 \omega_B^3}{2\omega^2} + \dots \,, 
\end{equation}
which  engenders Faraday rotation, and 
\begin{equation}
k_{\rm avg} \equiv  \frac{1}{2}(k_+ + k_-)
= {\omega} \left(1 - 
\frac{1}{2} \chi_0 \left(\frac{\omega_B}{\omega}\right)^2
- \frac{1}{8} \chi_0^2 \left(\frac{\omega_B}{\omega}\right)^2 + \dots \right)\,,
\end{equation}
which engenders time delay. Unlike the familiar gyroelectric Faraday effect, in which
electric charges are present, 
both the frequency dependence of the rotation 
and of the time delay are {\it trivial} in leading order in small quantities. 
At this order, the rotation 
angle, after transit through a length $l$, is 
\begin{equation}
\phi_0 = 
\frac{g\mu_M}{2} \int_0^l {\cal M}_0(x) dx \,, 
\label{rotang}
\end{equation} 
where ${\cal M}_0 = n_M \mu {\cal P}$ 
in a medium of spins of mass $M$, number density $n_M$,
and polarization ${\cal P}$. We note that the rotation angle is a signed quantity and can
tend to cancel if both particles and antiparticles are present. 

Terrestrial studies are tenable in this case 
because (i) 
we can apply a strong magnetic field of known strength, 
(ii) Faraday rotation accrues coherently under
momentum reversal, (iii) measurements of very small rotation angles are
possible~\cite{budker}, and finally (iv) entry into the magnetic field itself 
acts as a longitudinal Stern-Gerlach device~\cite{greene}. 
This last implies that we need not 
rely on any primordial polarization to detect an effect; rather, entry into a magnetic
field region itself acts as a spin filter device. This technique is used to polarize ultra-cold
neutrons (UCNs) with near 100\% efficiency in the UCNA experiment at Los Alamos~\cite{ucnA}.
The ``wrong'' (higher energy) spin state cannot 
enter the magnetic field region if it has a sufficiently low kinetic energy. 
Since the magnetization is determined by energy considerations, it is unaltered upon the 
replacement of particle with antiparticle and 
thus by $\mu \to - \mu$ under the $CPT$ theorem. However, under 
this replacement, the RHS of Eq.~(\ref{rotang}) changes sign. Faraday rotation probes
the properties of the medium; it is not a single-particle probe. If the 
particle-antiparticle symmetry were perfect, the rotation angle would vanish. 
One can also establish a dark-matter magnetic moment through 
experiments which search for anomalous 
recoils from spin-dependent scattering, though these studies 
are insensitive to the sign of the magnetic moment. 
Presuming sensitivity to comparable magnetic moments and masses, 
such studies and Faraday rotation studies 
are complementary. 
In contradistinction to scattering experiments, 
the Faraday effect can be used to discover whether 
an {\it asymmetry} in the dark sector is indeed present.

\section{Summary}

The preponderance of matter is unknown, and we can probe its
nature via its interactions with light. 
The discovery of dispersive effects in the speed of light
in propagation from distant GRBs at large redshifts 
would signal the presence of dark matter. 
The discovery of a non-zero millicharge would demonstrate that dark matter
is stable by dint of an internal $U(1)$ symmetry.
Studies of dispersive effects at optical energies and beyond
can constrain ``wimpless'' models and more.

We have also considered the possibility of observing a dark-matter candidate
particle with a non-zero magnetic moment through
the gyromagnetic Faraday effect. 
A non-zero Faraday rotation angle would signal that dark matter
possesses a particle-antiparticle asymmetry --- a unique insight.

\end{document}